\documentclass[a4paper,fleqn,usenatbib]{mnras}

\usepackage{mathptmx}
\usepackage{subfig}

\usepackage[T1]{fontenc}
\usepackage{ae,aecompl}

\usepackage{graphicx}	
\usepackage{amsmath}	
\usepackage{amssymb}	

\title[RGZ source classification with deep learning]{Radio Galaxy Zoo: Compact and extended radio source classification with deep learning}

\author[V. Lukic et al.]{
V. Lukic,$^{1}$\thanks{E-mail: vesna.lukic@hs.uni-hamburg.de}
M. Br\"uggen,$^{1}$\thanks{E-mail: mbrueggen@hs.uni-hamburg.de}
J.K. Banfield,$^{2,3}$
O.I. Wong,$^{4,3}$
L. Rudnick,$^{6}$
\newauthor R.P. Norris,$^{5,9}$
B. Simmons$^{7,8}$
\\
$^{1}$ Hamburger Sternwarte, University of Hamburg, Gojenbergsweg 112, Hamburg 21029, Germany \\
$^{2}$ Research School of Astronomy \& Astrophysics, Australian National University, Canberra, ACT 2611, Australia \\
$^{3}$ ARC Centre of Excellence for All-Sky Astrophysics (CAASTRO) \\
$^{4}$ International Centre for Radio Astronomy Research-M468, The University of Western Australia, 35 Stirling Hwy, Crawley, \\ WA  6009, Australia \\
$^{5}$ Western Sydney University, Locked Bag 1797, Penrith South, NSW 1797, Australia \\
$^{6}$ University of Minnesota, 116 Church St. SE, Minneapolis, MN 55455, USA \\
$^{7}$ Oxford Astrophysics, Denys Wilkinson Building, Keble Road, Oxford OX1 3RH, UK \\
$^{8}$ Center for Astrophysics and Space Sciences, Department of Physics, University of California, San Diego, CA 92093, USA \\
$^{9}$ CSIRO Astronomy and Space Science, Australia Telescope National Facility, PO Box 76, Epping, NSW 1710, Australia
}

\date{Accepted XXX. Received YYY; in original form ZZZ}

\pubyear{2018}

\begin{document}
\label{firstpage}
\pagerange{\pageref{firstpage}--\pageref{lastpage}}
\maketitle

\begin{abstract}
Machine learning techniques have been increasingly useful in astronomical applications over the last few years, for example in the morphological classification of galaxies. 
Convolutional neural networks have proven to be highly effective in classifying objects in image data. The current work aims to establish when multiple components are present, in the astronomical context of synthesis imaging observations of radio sources. To this effect, we design a convolutional neural network to differentiate between different morphology classes using sources from the Radio Galaxy Zoo (RGZ) citizen science project. In this first step, we focus on exploring the factors that affect the performance of such neural networks, such as the amount of training data, number and nature of layers and the hyperparameters. We begin with a simple experiment in which we only differentiate between two extreme morphologies, using compact and multiple component extended sources. We found that a three convolutional layer architecture yielded very good results, achieving a classification accuracy of 97.4\% on a test data set. The same architecture was then tested on a four-class problem where we let the network classify sources into compact and three classes of extended sources, achieving a test achieving a test accuracy of 93.5\%. The best-performing convolutional neural network setup has been verified against RGZ Data Release 1 where a final test accuracy of 94.8\% was obtained, using both original and augmented images. The use of sigma clipping does not offer a significant benefit overall, except in cases with a small number of training images.
\end{abstract}




\section{Introduction}

Extragalactic radio sources are among the most unusual and powerful objects in the universe. With sizes sometimes larger than a megaparsec, they have radio luminosities that are typically 100 times those of star-forming galaxies for example \citep{Van_Velzen_2012}, and display a wide range of morphologies. A new generation of wide-field radio interferometers are undertaking efforts to survey the entire radio sky to unprecedented depths making manual classification of sources an impossible task. Among the current and upcoming radio surveys that will detect such high numbers of radio sources are the LOw Frequency ARray 
(LOFAR\footnote{http://www.lofar.org}) surveys, Evolutionary Map of the Universe, the largest of such surveys in the foreseeable future \citep{Norris_2011}, VLA Sky Survey (VLASS\footnote{https://science.nrao.edu/science/surveys/vlass}) and surveys planned with the Square Kilometre Array (SKA\footnote{https://www.skatelescope.org}). The SKA alone will discover up to 500 million sources to a sensitivity of 2 $\mu$Jy/beam rms \citep{Prandoni_Seymour2015}. Radio interferometry data often display high levels of noise and artefacts \citep{Yatawatta_2008}, which presents additional challenges to any method of obtaining information from the data, such as extracting sources, detecting extended emission or detecting features through deep learning.

Machine learning techniques have been increasingly employed in data-rich areas of science. They have been used in high-energy physics, for example
in inferring whether the substructure of an observed jet produced as a result of a high-energy collision is due to a low-mass single particle or due to multiple decay 
objects \citep{Baldi_etal2016}. Some examples in astronomy are the detection of \textquoteleft weird' galaxies using Random Forests on Sloan data \citep{Baron_Poznanski2017}, Gravity Spy \citep{Zevin_2017} for LIGO detections, optimizing the performance and probability distribution function of photo-z estimation \citep{Sadeh_etal2016}, differentiating between real vs fake transients in difference imaging using artificial neural networks, random forests and boosted decision trees \citep{Wright_etal2015} and using convolutional neural networks in identifying strong lenses in imaging data \citep{Jacobs_etal2017}. 

Traditional machine learning approaches require features to be extracted from the data before being input into the classifier. Convolutional neural networks, a more recent machine learning method falling within the realm of deep learning, is able to perform automatic feature extraction. These suffer less information loss compared to the traditional machine learning approaches, and are more suited to high-dimensional datasets\citep{LeCun_Bengio_Hinton2015}. These are based on neural networks that contain more than one hidden layer \citep{Nielsen2015}. Each layer extracts increasingly complex features in the data before performing a classification or regression task. The raw data can be input into the network, therefore minimal to no feature engineering is required \citep{LeCun_etal1989}, and the network learns to extract the features through training. However, it should still be noted that convolutional neural networks do not always capture the data features.

The classification of optical galaxy morphologies is based on a few simple rules that makes it suitable for machine learning. It also lends itself to citizen science, where
these rules can be taught to non-experts. The Kaggle Galaxy Zoo \citep{Willett_etal2013} was a competition where the aim was to predict the probability distribution of 
the responses of citizen scientists about a galaxy's morphology using optical galaxy image data, and the winning solution used convolutional neural networks \citep{Dieleman_2015}. 

The convolutional neural network approach has only very recently started to be applied to radio galaxy images. One example has been in using convolutional neural networks to infer the presence of a black hole in a radio galaxy \citep{Alger_2016}. Another example is in a recently published paper by \citet{Aniyan_Thorat2017}, where the authors present their results on classifying radio galaxy images using convolutional neural networks into the classes of Fanaroff \& Riley Type 1 or 2 (FRI/ FRII) \citep{Fanaroff_1974} and bent-tailed radio galaxies using a few hundred original images in each class and producing a highly augmented dataset. They use a fusion classifier to combine the results of the three groups because poor results were achieved when attempting to do the three all together. Despite obtaining classification accuracies of above 90\% on the FRI and FRII classes, the authors have commented on issues with regards to overfitting due to having few representative samples in each class prior to augmentation, resulting in a small feature space and the fact that the network was highly sensitive to the preprocessing done to the images. 

In the case that outputs or labels are not provided alongside the input data to train on, one can use unsupervised machine learning techniques. In regards to machine learning with radio galaxy images, one method uses an unsupervised learning approach involving Kohonen maps (Parallelized rotation/flipping INvariant Kohonen maps, abbreviated to PINK) to construct prototypes of radio galaxy morphologies \citep{Polsterer_etal2016}. 

There are also automated methods that can help to generate labels, therefore the task becomes a supervised machine learning problem. In the astronomical context for example, there are source finding tools that can provide structure to data, and one such tool is PyBDSF \citep{Rafferty_Mohan2016}. This is the approach taken in the current work to provide the training labels.

The current work initially aims to classify radio galaxy morphologies into two very distinct classes, consisting of compact sources in one class and multiple-component extended sources in another class using convolutional neural networks. This setup we call the two-class problem. Once an optimal setup of parameters is found, we will test how it will work for the four-class problem of classifying into compact, single-component extended, two-component extended and multiple-component extended sources.

A compact source is an unresolved single component or point source, and an extended source is a resolved source, having at least one component. The detection of point sources is important as they are used for calibration purposes and they are also easier to match to their host galaxy. Making a proper census of unresolved sources is important for mapping out phase calibrators for radio interferometry \citep{Jackson_2016}. Although there are more conventional techniques to detect point sources, deep learning provides an alternative approach.

The Lasagne 0.2.dev1 library\footnote{https://lasagne.readthedocs.io/en/latest/} is used to build a deep neural network to differentiate between different classes of images of radio galaxy data. We compare the classifier metrics obtained on test samples, between the different models.

This paper is outlined as follows: Section~\ref{sec:dnn_theory} covers some basic theory about neural networks, and the advantages of using 
deep neural networks. In  Section~\ref{sec:methods} we discuss the data provided from Radio Galaxy Zoo, the minor pre-processing steps, and the use of algorithms to help select an image dataset consisting of compact and extended sources.  Section~\ref{sec:results_2} explores the two-class problem of distinguishing between compact and multiple-component extended sources. It documents the parameters and classifier metrics used. Section~\ref{sec:results_3} applies the optimal setup and parameters that were identified in Section~\ref{sec:results_2} to the four class problem of classifying between compact and three classes of extended sources. The best-performing setup is also tested to see how well it replicates the findings in Data Release 1 (DR1; Wong et al. in preparation) of the citizen science project Radio Galaxy Zoo. 

\section{Deep neural networks}
\label{sec:dnn_theory}

Neural networks can be used to perform classifications of data. If the input data is in the form of pixels of an image, along with corresponding labels for the image, this
information is fed into the input layer of the network \citep{Nielsen2015}. Neural networks are initialised with a set of weights and biases in the hidden layers \citep{Bishop_1995}. The data is propagated through the network and the output layer computes a prediction.  An error is calculated at the output layer using a cost or loss function, which is based on the difference between the true output and the predicted output \citep{LeCun_etal1998_2}. This error is back-propagated through the network, and the network adjusts the weights and biases to reduce the error \citep{Rumelhart_etal1986}. These steps are iterated a number of times until the cost function is minimised. This is known as training a neural network.

In feed forward neural networks, the nodes in the hidden layers are fully connected to the nodes in the adjacent layers. Therefore, the deeper the network becomes, the more
computationally intensive and time consuming it is to train, and often leads to the vanishing gradient problem \citep{Hochreiter_1991}. Convolutional neural networks have been shown to work much more efficiently in high-dimensional data such as image data \citep{Krizhevsky_etal2012} and although they still suffer from the vanishing gradient problem, one can lessen the impact by proper initialisation of the weights and biases, choosing an appropriate activation function and by doing layer wise pre-training. Such networks employ a number of filters of a certain size, as specified by the user. The receptive field is also referred to as the filter size. The filters are initialised with weights and biases from some distribution, and are connected to a small spatial portion of the input data. Features of the input data are learned through training. In image data, one can achieve a dramatic reduction in the number of parameters through parameter sharing, under the assumption of translational invariance. For example, if one feature is useful to compute at a particular spatial position, it should also be useful to compute at a different spatial position. Parameter sharing is achieved through the use of filters \citep{Karpathy2016}. One can reduce the computational complexity through data reduction with the use of pooling, in essence a subsampling method. There are several methods of implementing pooling such as max pooling and average pooling \citep{Lee_2016}. The current work uses max pooling, where the maximum value within a window of the input feature map is chosen. The convolutional and pooling layers are stacked with the end result being a hierarchical extraction of features. These layers are usually followed by one or more fully-connected layers, before finishing at the output layer, where a prediction is given \citep{Karpathy2016}.

One problem that occurs with neural networks is overfitting, which is when the architecture and parameters in the neural network fail to generalise to a separate dataset extracted from the same source, that has not been trained on. In this case, the model captures the noise in the data rather than the underlying signal, or there are real features in the training set that may be peculiar to individual sources but not common to the class as a whole. Overfitting is evident if the validation error is higher than the training error. To reduce the effect of overfitting, one can use image augmentation to artificially generate more images from the original data \citep{Krizhevsky_etal2012}. Another method is to use dropout in the dense or fully-connected layers, where a certain proportion of connections to nodes in adjacent layers are dropped to stop the network relying on the presence of particular neurons, hence it is made to be more robust \citep{Srivastava_etal2014}. Although early stopping is recommended to address the behaviour exhibited by deep neural networks trained on noise, defined as the memorization effect by \citet{Arpit_etal2017}, they find that such networks trained with Stochastic Gradient Descent learn patterns before memorizing, even in the presence of noise examples. 

\section{Methods}
\label{sec:methods}

We utilise the radio galaxy images from the Radio Galaxy Zoo project \citep{Banfield_etal2015}, which uses 1.4 GHz radio galaxy images from the Faint Images of the Radio Sky at Twenty Centimeters (FIRST). The original FIRST data reached a $1\sigma$ noise level of $150 \mu$Jy beam$^{-1}$ at 5$''$ resolution \citep{Becker_1995}. There are 206399 FITS files in total that contain single-channel image data. The size of the images is mainly (132,132) pixels resampled to a pixel size of 1.37$''$.

\subsection{Pre-processing}

The pixel values, representing brightness in mJy/beam were normalised by dividing by 255 such that the values are contained within the [0,1] range. Any \textquoteleft NaN' pixel value was converted to 0. The images were cropped to (110,110) pixels in order to slightly reduce the amount of data fed into the neural network. We were reluctant to do any further cropping because some of the extended sources tended to be very close to the image boundaries, which is information we did not want to remove. These were the only pre-processing steps taken to the original data. Later on we explore the effect of sigma clipping\footnote{http://docs.astropy.org/en/stable/api/astropy.stats.sigma\_clip.html} using a standard deviation of 3 to remove the background noise. This involves calculating the median (m) and standard deviation ($\sigma$) of the pixel values, and removing any value above $m+3\sigma$ and below $m-3\sigma$. However, deep neural networks should be able to account for the noise in the data without performing additional background noise removal. No procedure has been performed to remove artefacts in the data. As strong sidelobe emission is observed more often in the synthesis imaging of compact radio sources, sidelobe artefacts are expected to be minimal in RGZ and similarly so, for the purposes of this paper. \citet{Banfield_etal2015} added 5\% of the total sources as compact radio sources thus resulting in a smaller number where the sidelobe pattern could pose an issue.  Therefore, we do not expect large numbers of artefacts in the images to be misidentified as radio sources or components to cause an issue with our method. RGZ has a biased selection towards extended sources from the FIRST catalogue.

In order to provide an estimate of the presence of artefacts, we considered the sources in the two-component extended class from the four-class problem and found that 18 out of 11939 sources (0.15\%) contained one component having a total flux that was at least 50 times that of the other component.

\subsection{PyBDSF}

PyBDSF (the Python Blob Detector and Source Finder, formerly PyBDSM) by \citet{Mohan_Rafferty2015} is a tool designed to decompose radio interferometry images into sources composed of a set of Gaussians, shapelets, or wavelets. For the purposes of the current work, we assume that each image is of a single source or radio galaxy. Therefore, PyBDSF will detect the components belonging to the source.

In order to provide some initial structure to the data, we used the default settings of the PyBDSF version 1.8.11 \textquoteleft process\_image' task to help count the number of components in each image. The default settings include using 5-sigma for the pixel threshold and 3-sigma for island boundaries. The number of output lines in the resulting .srl file from running PyBDSF provides the user with the number of components that were able to be fit, using Gaussian fitting. The images were all initially run through PyBDSF. Out of the original 206399 images, 30945 produced an error, either due to the image having all blanked pixel values, presenting as NaNs (94.6\%), or there were no components detected in the image (5.4\%). 175454 images were successfully able to be processed by PyBDSF, and produced source list (srl) files that contained information about each detected source. In the successfully processed images, 99.7\% contained an NaN pixel percentage in the range between 0 and 10\%. The highest percentage of NaN pixel values was 93.2\% and the median was 1.9\%. The NaN values occur only along the edges of the images and are due to observations at the edges of fields. Table~\ref{tab:PyBDSF_number_components} lists the number of components detected in each image by PyBDSF, showing the results up to eleven components.

\begin{table}
	\centering
	\caption{The number of components detected by PyBDSF including how many of these sources there are, for up to 11 components.}
	\label{tab:PyBDSF_number_components}
	\begin{tabular}{lr}
		\hline
		PyBDSF number of components & Number of sources \\
		\hline
		1 & 63051 \\
		2 & 66589 \\
		3 & 29482 \\
		4 & 10437 \\
		5 & 3517 \\
		6 & 1136 \\
		7 & 510 \\
		8 & 264 \\
		9 & 163 \\
		10 & 79 \\
		11 & 48 \\
		\hline
	\end{tabular}
\end{table}

There are sometimes discrepancies between the number of components that PyBDSF had detected and how many there visually appeared to be in the image, therefore PyBDSF does not always perform as a human would in counting the number of components in the image. These inconsistencies remained even if the grouping parameters were altered. It was found that the number of inconsistencies detected increased with the number of components in the image. 

\subsection{Image augmentation}

The classification accuracy of deep neural networks increases with the size of the training set. It is possible to generate more images through label-preserving transformations such as horizontal, vertical translation and rotations \citep{Krizhevsky_etal2012}. This method is called augmentation and reduces the amount of overfitting to the data. It can also improve performance in imbalanced class problems \citep{Wong_etal2016}.

We augmented our images using translation, rotation and flips but not skewing or shearing the data since such transformations applied to compact sources can make them appear as having extended emission, which would render the label incorrect. The amount by which the images are translated is within the range of 0 to 22 pixels of the image width and height. Since no boundary conditions have been applied to the images, it is likely that 2.9\% of images in the two-class problem and 1.0\% in the four-class problem are likely to have components that have been shifted out of the image. The images are rotated by any random angle between 0 and 360 degrees. The Keras\footnote{https://keras.io/preprocessing/image/} package 2.0.3 was used to produce the augmented images. Keras is a high-level neural networks API, developed with the aim of enabling fast experimentation. It is written in the Python language and able to be run on top of either TensorFlow \footnote{https://www.tensorflow.org} or Theano\footnote{http://deeplearning.net/software/theano/}.

\begin{figure}
    \includegraphics[width=\columnwidth]{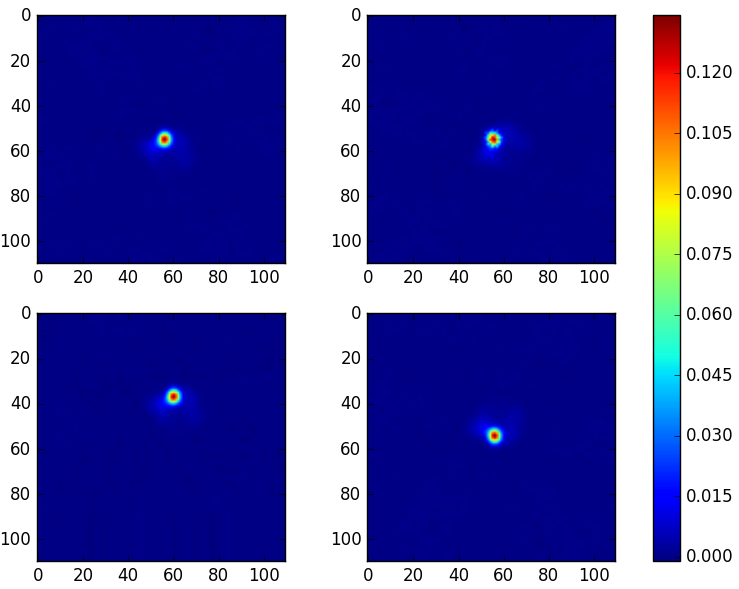}
    \caption{Examples of image augmentations with an extended source. The original image is shown on the top left. The transformations, from left to right,
    top to bottom are a random rotation, shift and flip. The size of the images is (110,110) pixels, with an angular resolution of 1.14$"$. The colour bar represents the normalised flux densities.}
    \label{fig:example_augment_rot_shift_flip}
\end{figure}

Fig.~\ref{fig:example_augment_rot_shift_flip} shows examples of rotation, shifting and flipping on a source with extended emission. The image is an example of how some extended sources that have a small amount of extended emission can look similar to compact sources, therefore presenting challenges for deep learning methods or other programs used to extract information from images.

\subsection{Deep learning algorithms}
\label{sec:DL_algorithms}
There are several deep learning implementations currently available for use. The present work uses Lasagne 0.2.dev1 \citep{Dieleman_et_al2015}, a 
lightweight library to build and train neural networks in Theano using the Python language. Python version 2.7.12 is used and the Theano version is 0.9.0dev2. Theano is a Python library that allows the user to define, optimize, and evaluate mathematical expressions involving multi-dimensional arrays efficiently. Some features of Theano include the ability to use Numpy arrays in Theano-compiled functions, the transparent use of a graphics processing unit (GPU), which enables data intensive computations to be accomplished much faster than on a CPU, and the ability to compute derivatives for functions with one
or many inputs. The python library Lasagne is built on top of Theano, but leaves the Theano symbolic variables transparent, so they can be easily altered 
to make changes to the model or learning procedure, as desired. This provides the user with more flexibility compared to that of other libraries.

Several network models built using the Lasagne library have been trained, in order to see the setup of parameters that results in the optimal test classification accuracy. The learning rate was set to $0.001$ at the beginning and reduced by a factor of 10 at four points during training. 1000 training epochs in total were used for all the models shown. The network parameters at the 1000$^{th}$ epoch were chosen for the final validation of the results. Training was stopped at this time because the training and validation losses appeared to reach their minimum and only fluctuated around this value, without the validation loss becoming higher than the training loss in the attempt to avoid overfitting, unless otherwise stated.

A simple manual tuning strategy was used to optimise the hyper-parameters, that involved experimenting with batch sizes of 8, 16 and 32 against different chunk sizes and learning rates. A batch size of 8 was found to give optimal results. The batch Stochastic Gradient Descent method \citep{Bottou_1998} was used, where the gradient is computed using the input dataset with a specified batch size, rather than using a single sample. The momentum update method used was Nesterov, with a momentum of 0.9 and a weight decay of 0. The Nesterov momentum update evaluates the gradient at the future rather than current position, resulting in better informed momentum updates and hence improved performance \citep{Sutskever_2013}. The validation step is done every 10 training epochs. 
The networks were trained on a single NVIDIA Tesla K20m GPU, with CUDA version 8.0.61. The categorical cross-entropy\footnote{http://lasagne.readthedocs.io/en/latest/modules/objectives.\\html\#lasagne.objectives.categorical\_crossentropy} cost function was used, which has the following form:

\begin{equation}
\label{eq:categorical_cross_entropy}
L_i= - \sum_j t_{i,j}\log(p_{i,j}) \,,
\end{equation}

\noindent where $i,j$ denotes the classes and observations respectively, $t_{i,j}$ represents the targets and $p_{i,j}$ represents the predictions.
Equation (\ref{eq:categorical_cross_entropy}) is used for predictions falling in the range (0,1) such as the softmax output of a neural 
network. The outputs of the softmax function represent the probabilities that the images belong to the given classes, and add up to 1. The predictions are clipped to be between $10^{-7}$ and $1-10^{-7}$ in order to make sure that they fall within the (0,1) range. There are over 1.6M parameters to train in total.

At the conclusion of training, the predictions at the final layer of the network are rounded to 0 or 1. In the two-class problem, the output [1,0] represents a compact source and [0,1] represents a multiple-component extended source. Training, validation and test classification accuracies are calculated using the proportion of rounded predictions that matched the labels. The image and label data have had the rows shuffled at two stages to make sure that there was no sampling bias when choosing the training, validation and test sets. A dropout of 50\% has been applied to the dense layers. The ReLU activation function \citep{Glorot_Bordes_Bengio2011} was used in the convolutional layers. The ReLU function is max(0,x), therefore only positive inputs are sent forward, and the negative ones are set to 0. This makes the network more sparse, therefore more efficient. Since the output is linear only in parts of the network, this ensures that the gradients flow across the active neurons, hence avoiding the vanishing gradient problem. The PReLU activation function \citep{He_etal2015}, which uses a negative linear function with a coefficient to control the negative part of the function was also tried, however it resulted in slightly worse accuracies compared to using the ReLU. In the dense layers, the identity activation function was used. The weights were initialised with the Uniform Glorot distribution \citep{Glorot_Bengio2010}, which has the following form when the ReLU activation function is used:

\begin{equation}
\label{eq:Glorot_uniform}
\sigma=\sqrt{\frac{2}{(n_{1}+n_{2})\cdot f}} \hspace{2mm} \,,
\end{equation}

\noindent where $n_1$ and $n_2$ is the number of connections coming in and out of the layer respectively, and $f$ is the receptive field size. The biases were initialised with the constant 0. 

In section~\ref{sec:Layers}, we explore the effect of varying the number of convolutional layers. Section~\ref{sec:Augmented} investigates the effect of adding augmented data for varying chunk sizes, and section~\ref{sec:Subset} explores the effect of using only a subset of the original provided images. The chunk size refers to the number of data examples per iteration and should be divisible by the batch size for optimal performance.

\subsection{Selection of sources for two-class classification}

In a first step, we applied a deep learning approach to two very distinct classes of radio sources: compact sources and multiple-component extended sources. Once this setup is optimised, we consider classification involving four classes.

In the current work, we define our sample of compact sources from the images where PyBDSF detected a single component, and additionally using Equation~(\ref{eq:theta_ratio}) from \citet{Kimball_Ivezic2008} as follows:

\begin{equation}
\label{eq:theta_ratio}
\theta=\Big(\frac{F_{\rm int}}{F_{\rm peak}}\Big)^\frac{1}{2} ,
\end{equation}

\noindent where $F_{\rm int}$ and $F_{\rm peak}$ are the integrated and peak flux intensities, respectively. According to this definition, values of $\theta \sim 1$ are highly 
concentrated (unresolved), while components with larger $\theta$ are extended (resolved). \citet{Kimball_Ivezic2008} adopt $\theta \approx 1.06$ as the value separating 
resolved and unresolved components, where components above $\theta \approx 1.06$ are resolved. We therefore define our compact components as those having values $\theta < 1.06$. If there was only one compact component in the image, then it was classified as a \textquoteleft compact' source; there were 2682 such cases. The $F_{\rm int}$ and $F_{\rm peak}$ values were extracted from the provided FITS files, using the \textquoteleft imfit' function from CASA Version 4.7.2-REL. Several batches of samples assigned to the \textquoteleft compact' class were additionally examined visually to verify that they truly appeared to be compact sources.

The choice of multiple-component extended sources was taken from a random sample of 18000 images where PyBDSF had detected at least 3 components. This sample can include images of multi-component compact sources.

\begin{figure}
    \includegraphics[width=\columnwidth]{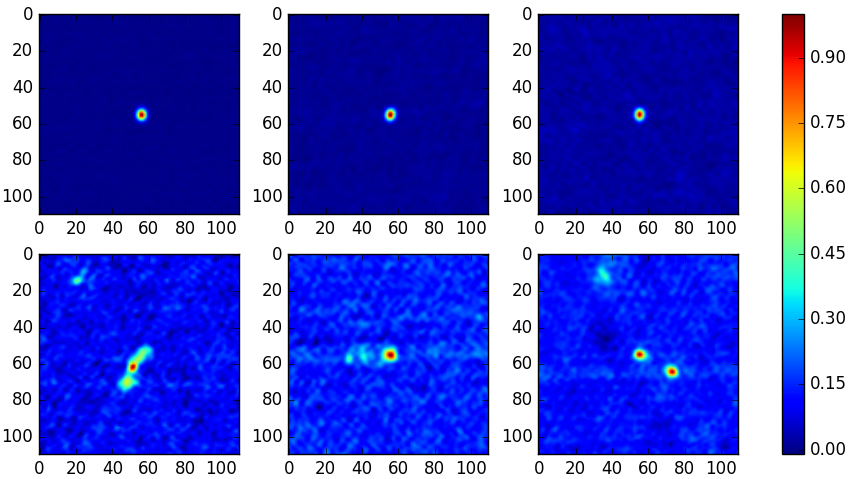}
    \caption{Examples of compact and multiple component extended source classifications that we initially aim to make our deep neural networks differentiate between. The top row of images represents compact sources, whereas the bottom row represents multiple component extended sources. The colour bar represents the normalised flux densities.}
    \label{fig:point_other_examples}
\end{figure}

Taking this sample of compact and extended sources, there are 20682 images all together that can be divided into a training, validation and test data set for the initial
deep learning approach. The number of images used for the two-class problem is summarised in Table \ref{tab:Number_summary}. Fig.~\ref{fig:point_other_examples} shows some typical examples of compact and multiple-component extended sources. It should be noted that there are many more examples of multiple extended sources compared to compact sources, however the compact sources display a very well defined morphology compared to the multiple extended sources, which can assume an almost infinite number of unique morphologies. 

In examining the images where PyBDSF had detected at least three components, it appears that some of the images contain superpositions, or have fewer than three components in the image. Upon closer inspection of a random sample of 250 images where PyBDSF has detected at least three sources, there were roughly 44\% that appeared to be superpositions or that visually had fewer than three components in the image. This means that a substantial number of images assigned to the multi-component class do not truly belong, however there is still a stark contrast in morphology compared to the sources chosen for the compact class, therefore it should not have an overly detrimental effect on the classification accuracies. We attempt to eliminate these contaminated images when choosing sources for the four-class problem.

\begin{table}
	\centering
	\caption{Summarising the number of images used for the two-class problem}
	\label{tab:Number_summary}
	\begin{tabular}{lrr}
		\hline
		Source/Image type & \# Original & \# Augmented \\
		\hline
		Compact & 2682 & 15558\\
		Multiple-extended & 18000 & 144633\\
		{\bf Total} & 20682 & 160191\\
		\hline
	\end{tabular}
\end{table}

\subsection{Selection of sources for four-class classification}

Assuming there is an optimal choice in hyper-parameters that results in a high classification accuracy for the two-class problem of distinguishing between compact and multiple-component extended sources, we also wanted to see how such a setup would be able to distinguish between sources belonging to four classes. We chose the images belonging to categories of compact sources, single component extended, two component extended and multiple component extended sources. The compact sources are the same ones as were used for the two-class problem, and the multiple-component extended sources are a subset of the ones used for the two-class problem. The single-component extended and two-component extended classes are the new classes, and the images belonging to them have not previously been used for the deep learning approach. 

The labels for the images were able to be generated with the help of the \textquoteleft S\_code'  output of PyBDSF. The S\_code quantity defines the component structure \citep{Mohan_Rafferty2015} and the output values are defined as such:

\begin{itemize}
 \item \textquoteleft S' = a single-Gaussian component that is the only component in the island
 \item \textquoteleft C' = a single-Gaussian component in an island with other components
 \item \textquoteleft M' = a multi-Gaussian component
\end{itemize}

\noindent The four classes are described below: 

\begin{itemize}
 \item Compact source: Sources where PyBDSF has detected one component and choosing sources as defined by Equation~(\ref{eq:theta_ratio}) from \citet{Kimball_Ivezic2008}. The same set of compact sources were used for the two-class problem.
 \item Single component extended source: Sources where PyBDSF has detected one component, and the S\_code quantity contains an \textquoteleft M' 
 (multi-Gaussian component).
 \item Two component extended sources: Sources where PyBDSF has detected two components, and the S\_code quantity contains an \textquoteleft M' 
 (multi-Gaussian component) for both components.
 \item At least three component extended sources: Sources where PyBDSF has detected at least three components. We started with the set of 18000 images as for the two-class problem, required that the S\_code quantity contains at least two \textquoteleft M's, and any number of \textquoteleft C's. Additionally, two blob-detection algorithms (logarithm of gaussian and difference of gaussian) were run using the scikit-image 0.17.1 package in Python\footnote{http://scikit-image.org/docs/dev/auto\_examples/\\features\_detection/plot\_blob.html}. The images were also all inspected visually in an attempt to ensure that each image contained at least three extended components that appeared to be part of the same source, rather than being superpositions of sources. After this, there were 577 images remaining. However upon cross-checking with several optical/IR images, more than 40\% of this subset of images still appeared to contain superpositions of components associated with more than one AGN. Therefore, although the classification successfully identifies multiple-component structures, they are contaminated by such superpositions in comparison with Radio Galaxy Zoo classifications.
\end{itemize}

\begin{figure}
    \centering
    \includegraphics[width=85mm,scale=0.9]{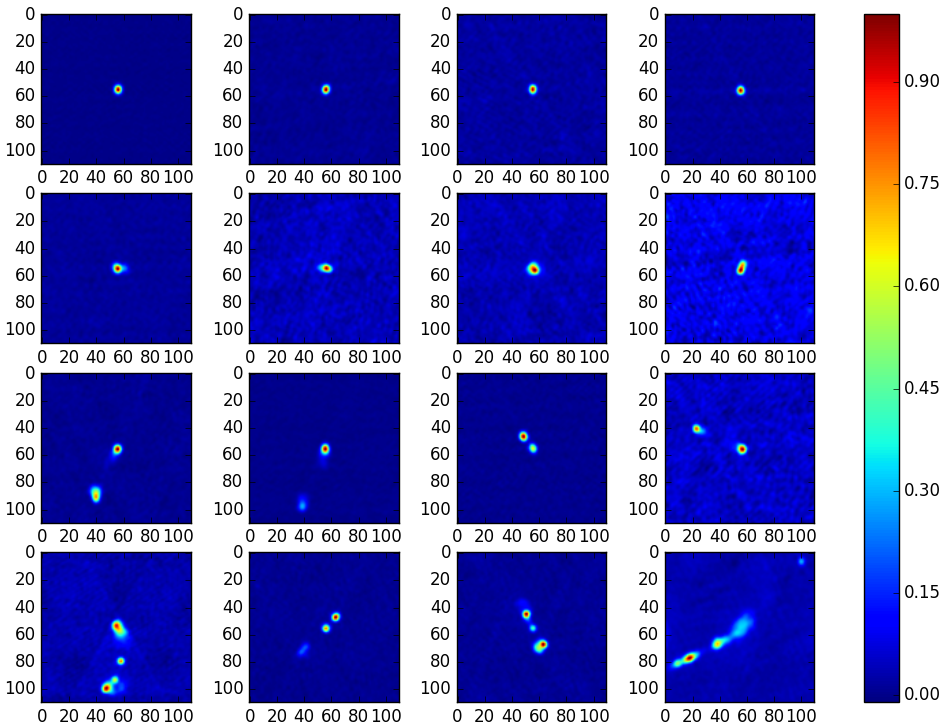}
    \caption{Examples of compact, single-extended, two-component extended and multiple-component extended sources, for a deep neural network to differentiate between. Top row: Compact
    sources. Second row: Single-component extended sources. Third row: Two-component extended sources. Fourth row: Multiple-component extended sources.}
    \label{fig:four_class_examples}
\end{figure}

\noindent The condition \textquoteleft S\_code=S' was not found to be useful in characterising components. Occasionally there was a source that appeared as though it should belong to another class, so a small level of label contamination must be accepted. The number of images used for the four-class problem is summarised in Table \ref{tab:Four_class_problem1} and Fig.~\ref{fig:four_class_examples} shows some example images for each of the four classes. The four-class classification scenario also contains an imbalance in the number of original images for each class, however this can be alleviated by augmenting the classes of data displaying richer morphologies more (single, two extended and multiple component extended sources), compared to the compact sources.

The existence of the remaining superpositions of sources in the multiple-component extended class in the training set means that the deep learning algorithm will not be able to make the distinction between images that contain superpositions, and images with components that are likely to be part of the same source. Even radio galaxy experts cannot always reach a consensus about these differences.

The fact that the compact and single-component extended sources all come from the set of images where PyBDSF has detected a single source mean that the deep learning algorithm is doing more than just learning the method by which PyBDSF counts components. It is also performing the source structure functions of PyBDSF, with the advantage that it uses solely image data to learn about the differences in morphology between compact sources and single component extended sources.
 
\begin{table}
	\centering
	\caption{Summarising the number of images used for the four-class problem.}
	\label{tab:Four_class_problem1}
	\begin{tabular}{lrr}
		\hline
		Source/Image type & \# Original & \# Augmented \\
		\hline
		Compact & 2682 & 15558\\
		Single-extended & 6735 & 43099\\
		Two-extended & 11939 & 35994\\
		Multiple-extended & 577 & 46381\\
		{\bf Total} & 21933 & 141032\\
		\hline
	\end{tabular}
\end{table}

\section{Results for two classes}
\label{sec:results_2}

Our first aim is to see how well a deep neural network is able to distinguish between two classes of data that are very morphologically distinct: compact sources and multiple component sources. There were 2682 compact and 18000 multiple-component extended sources, giving a total of 20682 images provided as input data for classification by the convolutional neural network designed in Lasagne. When the augmentation data is used as well, there are a total of 180873 images. The number of sources and augmented images used is summarised in Table \ref{tab:Number_summary}.

The results shown are the classifier metrics on the validation and test data sets. The extended source class is used as the positive class for the 
metrics, therefore a true positive (TP) is defined as when an extended source is predicted that is also labelled as an extended source. A false positive (FP) is defined when an extended source is predicted, but is labelled as a compact source. A false negative (FN) is defined when a point source is predicted, but is labelled as an extended source. The following four metrics are used to
evaluate the performance of the classifier:

\begin{itemize}
 \item $\textrm{Precision = TP/(TP+FP)}$
 \item $\textrm{Recall = TP/(TP+FN)}$
 \item $\textrm{F1 score =(2}$$\times$Precision$\times$Recall)/(Precision + Recall)
 \item $\textrm{Accuracy = (TP+TN)/(TP+FP+TN+FN)}$
\end{itemize}

\noindent where FP, FN and TN denotes false positives, false negatives and true negatives respectively. For the current task of classifying between extended and point sources, precision represents the classifier's ability to not classify point sources as extended sources. Recall evaluates the classifiers ability to not classify extended sources as point sources, hence provides an estimate of the sensitivity of the classifier, in whether it can correctly predict the labeled extended sources. It is worth noting that in the literature, precision is often called \textquotedblleft reliability" and recall is often called \textquotedblleft completeness" (e.g. \citet{Hopkins_etal2015}). The F1 score can be interpreted as the weighted average of precision and recall. The accuracy is the overall classification accuracy across the classes, how many correct predictions for the labeled point and extended sources were made overall. The F1 and accuracy scores tend to correlate highly. The precision, recall, F1 score and accuracy metrics were calculated for both the validation and test data sets to assess the performance of each deep neural network model. It should be noted that in machine learning theory, the precision scores are a better assessor of performance compared to accuracy in imbalanced dataset problems. However, we address the imbalance in our dataset through augmentation, therefore use the classification accuracy to assess the performance of the classifiers. The training and validation losses are also plotted as a function of epochs for several chosen models. 

In order to assess which models are significantly better than others, rather than arising as a result of random fluctuations, we use the root mean square error (RMSE) measure to quantify the scatter in the overall accuracies, for the original data. The RMSE is defined according to Equation \ref{eq:RMSE}.

\begin{equation}
\label{eq:RMSE}
RMSE=  \sqrt{\frac{1}{n} \sum_j^n (p_{i,j}- t_{i,j})^2} \,,
\end{equation}

\noindent where $i,j$ denotes the classes and observations respectively, $n$ is the total number of observations, $t_{i,j}$ represents the targets and $p_{i,j}$ represents the predictions. We consider any value beyond two times the RMSE value to be significant in terms of metrics. This error estimate is conservative in that it is a measure of the actual scatter, as opposed to the derived error in the mean accuracies.

\subsubsection{Effect of increasing convolutional layers}
\label{sec:Layers}

We first explore the effect of increasing the number of convolutional layers, in order to see the effect the model complexity has on obtaining better classification accuracies, without excessive overfitting.

\begin{figure}
    \centering
    \includegraphics[width=90mm,scale=0.9]{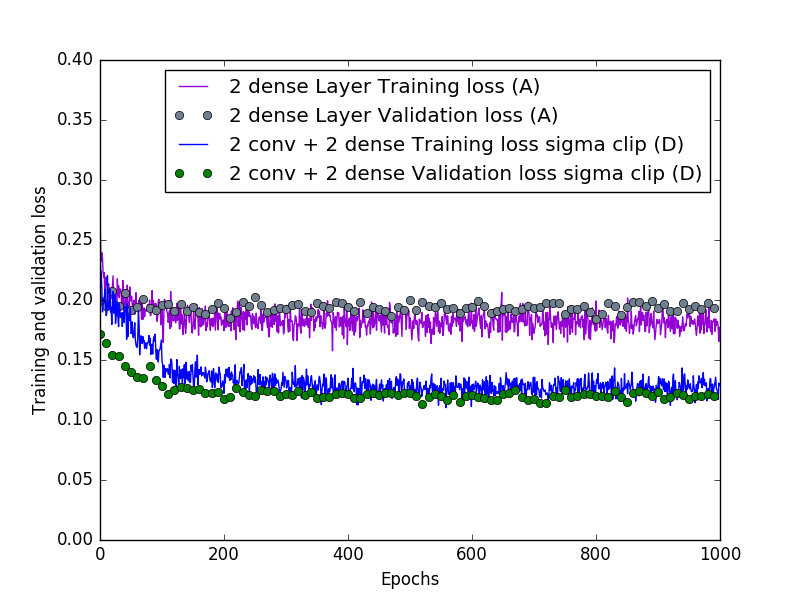}
    \caption{Plot of training and validation losses as a function of training epochs, for models A and D in Table~\ref{tab:Orig_models}. The higher training and validation losses
    are from using only 2 dense layers and no convolutional layers, which are the highest losses amongst the six models and consequently produced the lowest classification accuracies. Adding 2 convolutional layers produces lower training and validation losses, and therefore improved classification accuracies. The 2 convolutional and 2 dense layer architecture with sigma clipping was one of two models that performed the best out of all the models considered for this set of images.}
    \label{fig:Train_plots_combined}
\end{figure}

\begin{figure}
    \centering
    \includegraphics[width=90mm,scale=0.9]{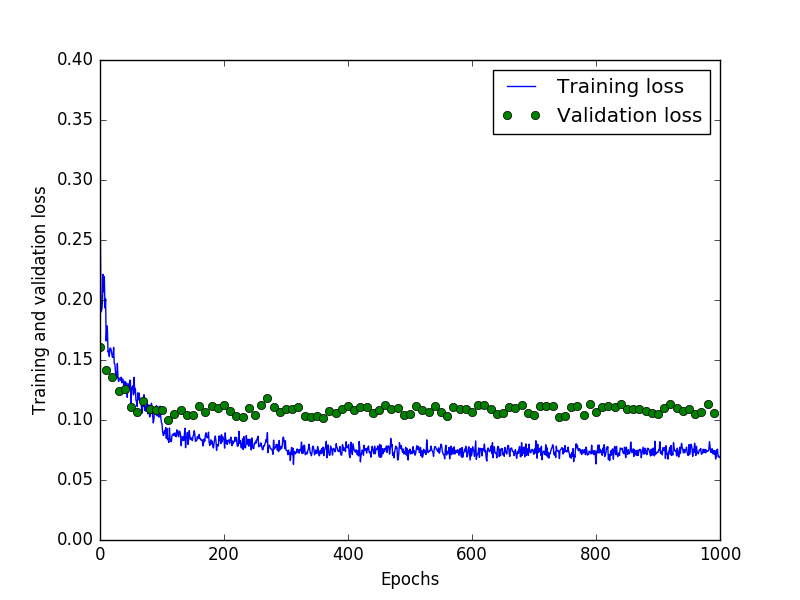}
    \caption{Plot of training and validation losses as a function of training epochs, for a three convolutional and two dense layer model with sigma clipping (model F). Despite this model achieving the highest test accuracy for the original set of images, overfitting is evident as the validation losses are higher than the training losses.}
    \label{fig:Sigma_overfit}
\end{figure}

The results in Tab.~\ref{tab:Results1} and Fig.~\ref{fig:Train_plots_combined} show the effect of adding an increasing number of layers to the network. Simply using two dense layers results in precision, recall and F1 scores above 0.95 and a test accuracy above 93\%. The addition of two adjacent convolutional layers and the use of sigma clipping produces a classification accuracy of 97.0\%. Taking into account the RMSE values to establish random fluctuations in accuracy, the model that is significantly better than all others is the three convolutional and two dense layer model with sigma clipping (model F), achieving the optimal accuracy of 97.5\%. However, this setup results in overfitting as shown in Fig.~\ref{fig:Sigma_overfit}, and hence we exclude this model. The next best-performing models that are significantly better than the others, without causing overfitting, are models D and E.

Using two adjacent convolutional layers followed by a pooling layer as opposed to using a single convolutional layer followed by a pooling layer reduces the number of parameters, given that the two filter sizes of the adjacent convolutional layers are smaller compared to using a single larger one \citep{Simonyan_Zisserman2015}. When putting a max pooling layer in between the first and second convolutional layer, it had 
a detrimental effect on the test accuracy, reducing it by almost 1\% which is significant given the RMSE values, and it took more training epochs to attain a smaller training loss (results not 
shown). The radio galaxy images with extended emission generally have structure that span across large portions of the image, yet it would 
increase the number of parameters by too much of a factor if a single convolutional layer with a very large receptive field, or filter size was used. Therefore, 
it is better to combine two adjacent convolutional layers that have smaller receptive fields.

The deep learning algorithm appears to be robust to the classes being imbalanced; there are approximately 9 times more examples of the multiple extended class images compared to the compact source images. However, the compact sources have a much more stable morphology, largely consisting of a source in the centre of the image, compared to the multiple component extended class images, which can be spread out all over the image.

Considering the results for the test data set and taking into account the RMSE values, the precision (reliability) values are on average significantly higher compared to the recall (completeness) values. This implies that the classifier is better at not classifying the multiple-component extended sources as point sources but is not as sensitive in identifying all the labeled multiple-component extended sources. The training losses begin at a low value of around 0.27 and quickly settle to their minimal value for a particular model by 200 epochs. A likely reason why the losses begin and remain low during training is because the classes contain images that are morphologically very different; one containing a single concentrated source in the centre of the image and the other generally containing multiple sources that are spread throughout the image.

The fact that a very substantial number of images belonging to the multiple-component extended class appear to contain superpositions or visually appear as though they contain fewer than three components probably does not hinder the classification accuracies significantly, since the contents of the images are very different between the two classes.

The memory requirements for a typical run using the three convolutional layer architecture is 1.87 GB, with a computational time of 192 minutes using a single NVIDIA Tesla K20m GPU, with CUDA version 8.0.61.

\begin{table}
	\centering
	\caption{The deep learning models that were explored.}
	\label{tab:Orig_models}
	\begin{tabular}{lcr}
	\hline
	Code & Model & \# Pooling layers \\
	\hline
	A & 2 dense & 0 \\
	B & 1 conv + 2 dense & 1 \\
	C & 2 conv + 2 dense & 1 \\
	D & 2 conv + 2 dense sigma clip & 1 \\
	E & 3 conv + 2 dense & 2 \\
	F & 3 conv + 2 dense sigma clip & 2 \\
	\hline
	\end{tabular}
\end{table}

\begin{table}
	\centering
	\caption{Effect of increasing the number of convolutional layers for the original images. The precision, recall, F1 score and accuracy values are shown for both the validation and test data 
	sets, calculated over 1000 training epochs. The validation set is used every 10 epochs, and the final trained parameters are used on the test data set after training
	is complete. 20682 images were used in total, with a chunk size of 6000, and the training samples make up 60\% of the total data.}
	\label{tab:Results1}
	\begin{tabular}{lccccr}
		\hline
		 & \textbf{Valid.} Prec. & Recall & F1 score & Accuracy & RMSE \\
		\hline
		A & 96.6\% & 95.6\% & 96.1\% & 93.3\% & 0.27 \\
		B & 97.9\% & 97.0\% & 97.4\% & 95.6\% & 0.22 \\
		C & 97.4\% & 97.5\% & 97.4\% & 95.6\% & 0.20 \\
		D & 98.2\% & 96.9\% & 97.5\% & 95.7\% & 0.21\\
		E & 98.6\% & 97.5\% & 98.0\% & 96.6\% & 0.19 \\
		F & 98.4\% & 97.5\% & 97.9\% & 96.4\% & 0.19 \\
		\hline
		& \textbf{Test.} Prec. & Recall & F1 score & Accuracy & RMSE \\
		\hline
		A & 97.4\% & 95.7\% & 96.6\% & 94.0\% & 0.26 \\
		B & 98.2\% & 96.3\% & 97.3\% & 95.3\% & 0.22 \\
		C & 97.7\% & 96.7\% & 97.2\% & 95.1\% & 0.21\\
		D & 98.1\% & 98.4\% & 98.3\% & 97.0\% & 0.19\\
		E & 98.2\% & 97.8\% & 98.0\% & 96.5\% & 0.21 \\
		F & 98.7\% & 98.3\% & 98.5\% & 97.5\% & 0.18 \\
		\hline
	\end{tabular}
\end{table}

\subsubsection{Effect of including augmented data}
\label{sec:Augmented}

\begin{figure}
    \centering
    \includegraphics[width=90mm,scale=0.9]{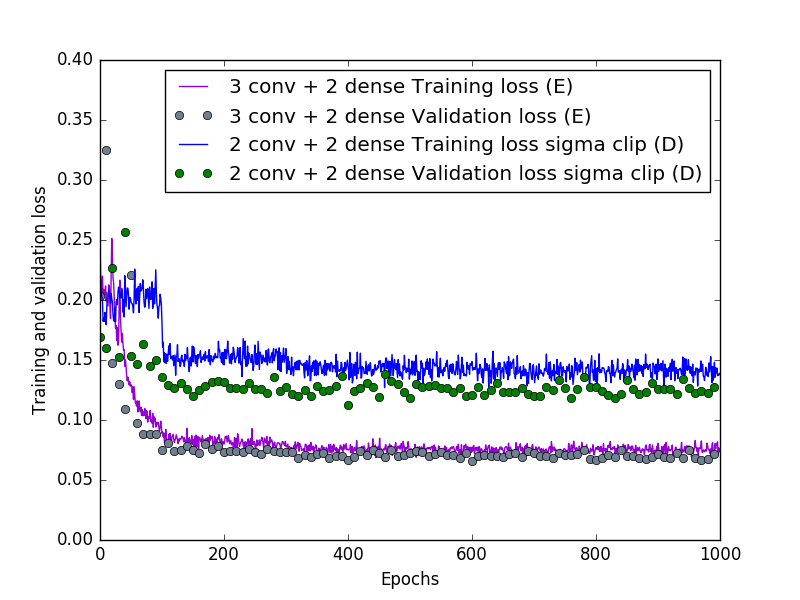}
    \caption{Plot of training and validation losses for the 2 conv + 2 dense layer with sigma clipping (model D) and 3 conv + 2 dense layer (model E), when using the original and augmented data. The training and validation losses are higher and fluctuate more for model D, and there is a greater difference between the training and validation losses compared to model E, despite achieving a similar test classification accuracy. Taking these factors into account, model E performs better overall.}
    \label{fig:Augmented_plots}
\end{figure}

Next we studied the effect of image augmentation on the classification accuracies. Table~\ref{tab:Results2} and Fig.~\ref{fig:Augmented_plots} shows that when the full set of augmented data is used in addition to the original images, it results in overall significantly improved F1 scores, validation and test accuracies, compared to when the original data is used. The use of the augmented images enables the choice of the larger chunk size leading to improved accuracy without causing the network to overfit, hence a chunk size of 20000 is used, compared to the previous size of 6000. The chunk size should be made as large as possible for a given set of data, since the more training examples are seen simultaneously, the more accurately the weights can be adjusted to produce a lower training loss. The best performing architecture with the original and augmented images is the three convolutional and two dense layer architecture, with no sigma clipping (model E). This setup achieves the highest observed test accuracy for the two-class problem of 97.4\%. Model D performs equally well in terms of overall accuracy when taking into account the RMSE values, however there is a greater difference in the training loss compared to the validation loss, as is evident in Fig.~\ref{fig:Augmented_plots}. Therefore, model E is the overall best-performing model, since the training and validation losses are closer together.

\begin{figure}
    \centering
    \includegraphics[width=50mm,scale=0.5]{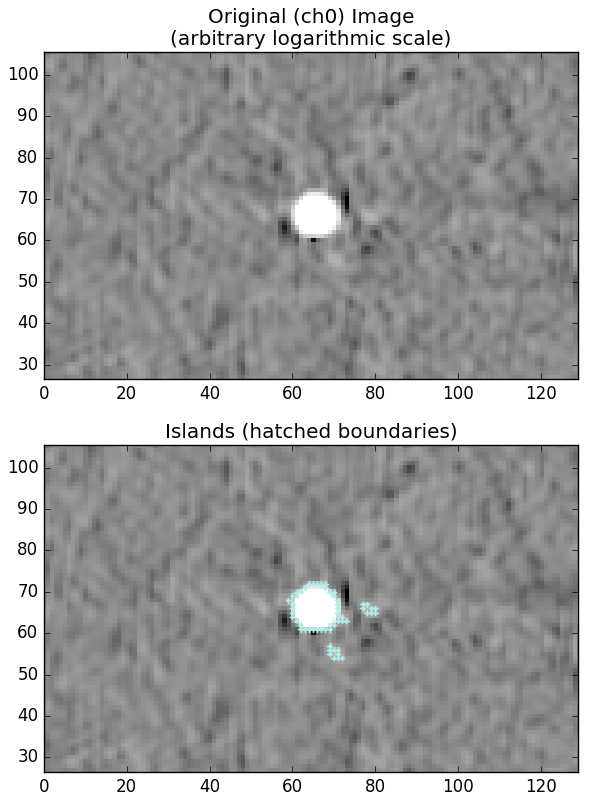}
    \caption{Example of an image where PyBDSF has detected 3 components, even though the image appears to be that of a point source.}
    \label{fig:3_component_image}
\end{figure}

\begin{figure}
    \includegraphics[width=\columnwidth]{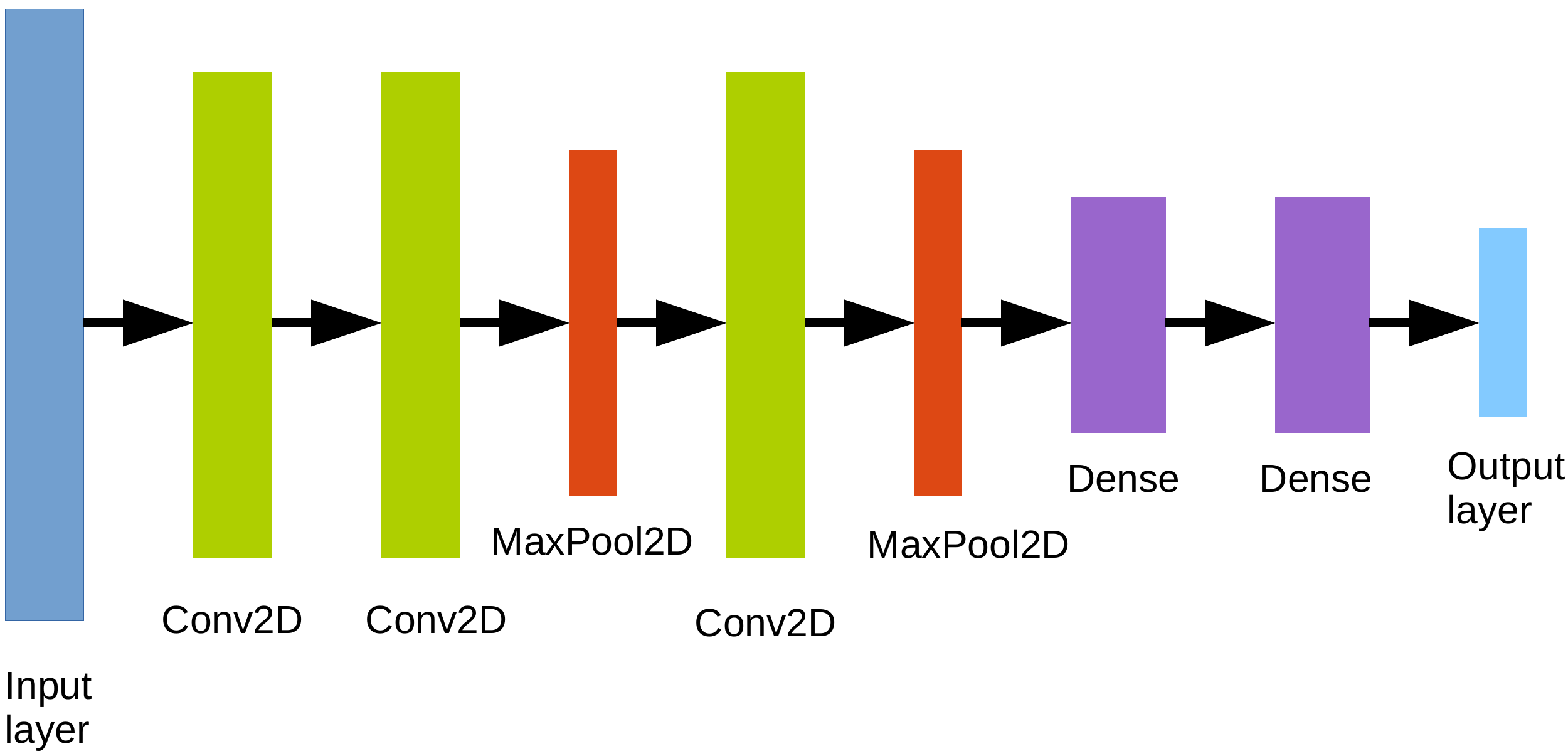}
    \caption{The 3 conv + 2 dense architecture, which constituted the best performing model. The colours are arbitrarily chosen to represent the different layers used.}
    \label{fig:Best_architecture}
\end{figure}

The most likely reason why a higher accuracy is unable to be achieved is that there is a small amount of label contamination, for example a few of the images in the multiple component extended class may look more like compact sources. This is due to PyBDSF detecting multiple components in an image, even though visually the image appears to only contain a compact source, as shown in Fig.~\ref{fig:3_component_image}. The three convolutional and two-dense layer architecture is shown in Fig.~\ref{fig:Best_architecture}, and the details of the layers with the number of parameters used are shown in Tab.~\ref{tab:Model_parameters}.

\begin{figure}
    \includegraphics[width=\columnwidth]{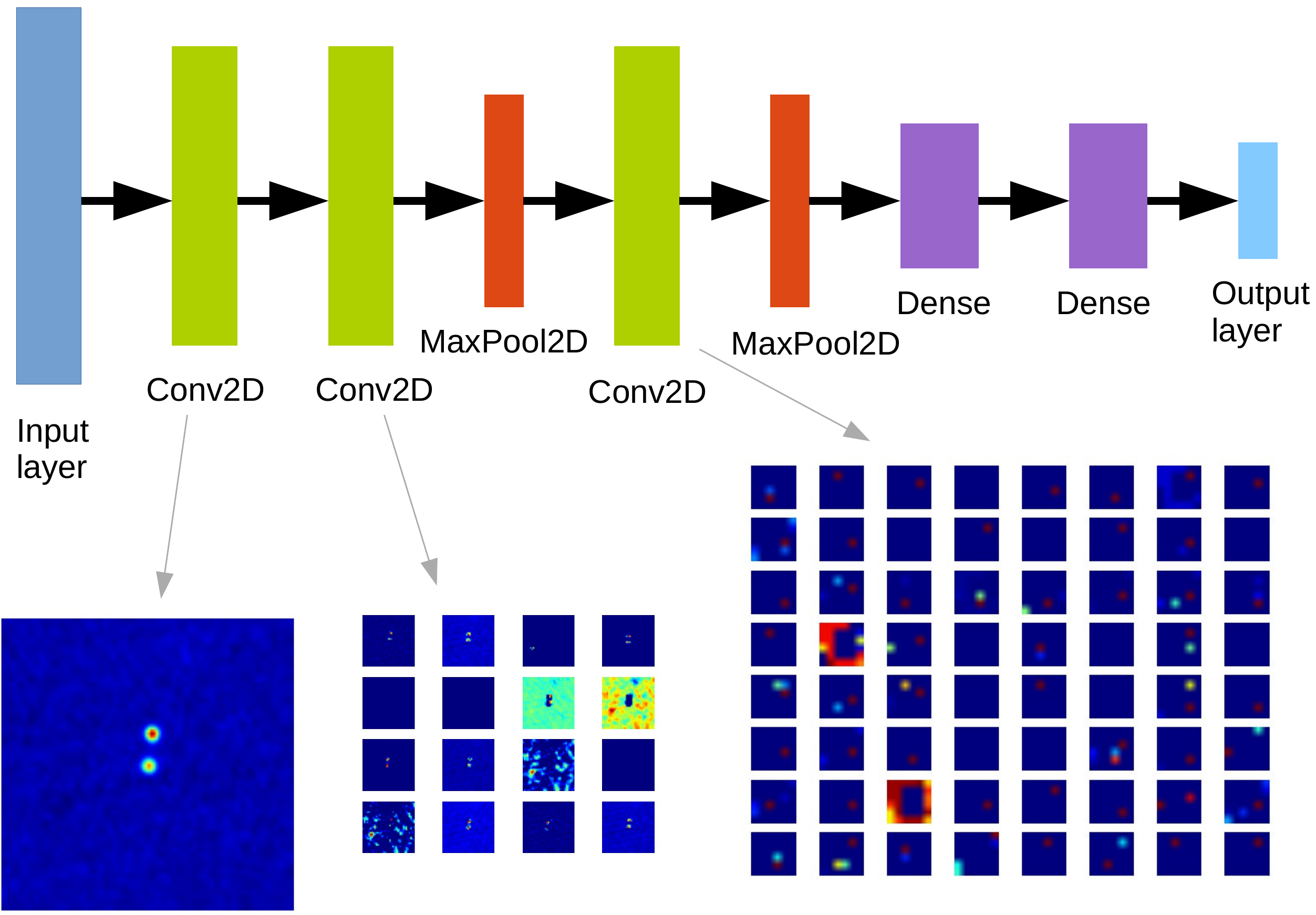}
    \caption{Showing the input image, first and third convolutional feature map activations at 100 epochs into training using the three convolutional and two dense layer architecture. The colours in the architecture are arbitrarily chosen to represent the different layers used.}
    \label{fig:feature_maps}
\end{figure}

Fig.~\ref{fig:feature_maps} shows the features that are learnt in the first and third convolutional layers for the three convolutional layer architecture, halfway through 
training at 500 epochs.

\begin{table}
	\centering
	\caption{Effect of using all augmented images in addition to original data. The precision, recall, F1 score and accuracy values are shown for both the validation and test data sets, calculated over 1000 training epochs. The validation set is used every 10 epochs, and the final trained parameters are used on the test data set after training is complete. 180873 images were used in total, with a chunk size of 20000, and the training samples make up 60\% of the total data}
	\label{tab:Results2}
	\begin{tabular}{lcccr}
		\hline
		& \textbf{Valid.} Precision & Recall & F1 & Accuracy \\
		\hline
		C & 97.0\% & 98.2\% & 97.6\% & 95.7\% \\
		D & 98.3\% & 98.3\% & 98.3\% & 96.9\% \\
		E & 98.8\% & 97.9\% & 98.4\% & 97.0\% \\
		F & 98.6\% & 97.8\% & 98.2\% & 96.8\% \\
		\hline
		& \textbf{Test} Precision & Recall & F1 & Accuracy \\
		\hline
		C & 96.6\% & 98.5\% & 97.6\% & 95.6\%  \\
		D & 98.7\% & 98.4\% & 98.5\% & 97.4\%\\
		E & 99.2\% & 97.9\% & 98.6\% & 97.4\% \\
		F & 98.7\% & 97.8\% & 98.2\% & 96.9\% \\
		\hline
	\end{tabular}
\end{table}

\begin{table}
	\centering
	\caption{Details of the layer parameters used for the best-performing model. The \# of parameters gives a cumulative sum at each layer. There
	are 1676914 trainable parameters in total.}
	\label{tab:Model_parameters}
	\begin{tabular}{lcccr}
		\hline
		Layer & Depth & Filter Size & Stride length &  \#Parameters \\
		\hline
		Conv2D & 16 & 8 & 3 & 1040 \\
		Conv2D & 32 & 7 & 2 & 26160 \\
		MaxPool2D & 32 & 3 & - & 26160 \\
		Conv2D & 64 & 2 & 1 & 34416 \\
		MaxPool2D & 64 & 2 & - & 34416 \\
		Dense & 1024 & - & - & 625264 \\
		Dense & 1024 & - & - & 1674864 \\
		Softmax & 2 & - & - & 1676914 \\
		\hline
	\end{tabular}
\end{table}

\subsubsection{Effect of using a subset of images}
\label{sec:Subset}

\begin{figure}
    \centering
    \includegraphics[width=90mm,scale=0.9]{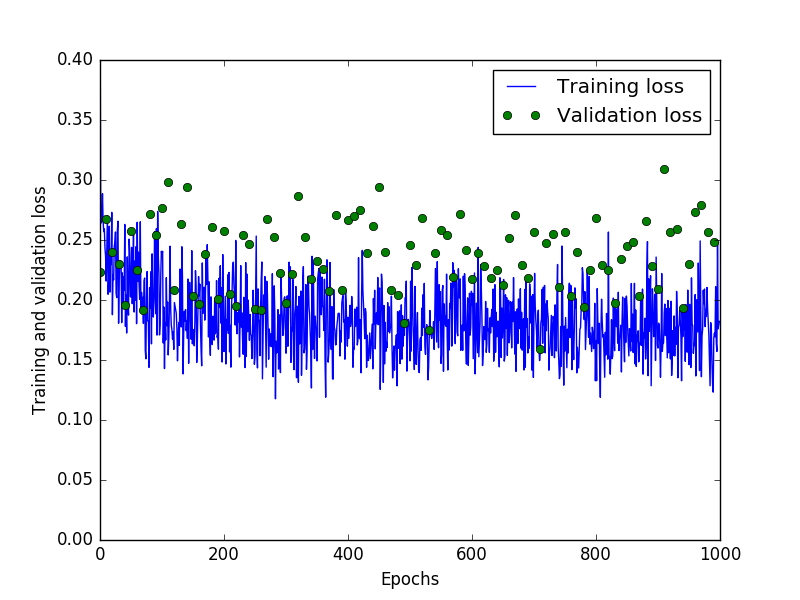}
    \caption{Training and validation losses when using only 1000 original and 1000 augmented images, when using the 2 convolutional and 2 dense layer setup with sigma clipping (model D). The training losses are around the same compared to when using the full set of 20682 images, and the fluctuations are greater. There is also some amount of overfitting.}
    \label{fig:1000_original}
\end{figure}

Next we explored the effect of using only a subset of images. Using a subset of the available images (1000 original and 1000 augmented) tended to significantly reduce the validation and test scores compared to when using the full set of original images, as shown in Tab.~\ref{tab:Results3}, caused greater fluctuations during training, and introduced a higher level of overfitting as shown in Fig.~\ref{fig:1000_original}. The larger fluctuations during training are most likely due to the algorithm not seeing as large a number of samples at a time compared to when the full set of images is used, hence the weights cannot be estimated as accurately for each subsequent training epoch. The validation and test accuracies however still remained above 90\%, with the exception of model F (three convolutional and two dense layer setup with sigma clipping.)

\begin{table}
	\centering
	\caption{Effect of using a subset of the original and augmented images. The precision, recall, F1 score and accuracy values are shown for both the validation and test data 
	sets, calculated over 1000 training epochs. The validation set is used every 10 epochs, and the final trained parameters are used on the test data set after training
	is complete. 1000 original and 1000 augmented images were used (2000 in total) with a chunk size of 400, and the training samples make up 60\% of the total data.}
	\label{tab:Results3}
	\begin{tabular}{lcccr}
		\hline
		& \textbf{Valid.} Precision & Recall & F1 & Accuracy \\
		\hline
		C & 94.1\% & 97.8\% & 95.9\% & 92.7\% \\
		D & 95.2\% & 96.4\% & 95.8\% & 92.6\% \\
		E & 95.3\% & 96.2\% & 95.7\% & 92.4\% \\
		F & 95.3\% & 96.3\% & 95.8\% & 92.5\% \\
		\hline
		& \textbf{Test} Precision & Recall & F1 & Accuracy \\
		\hline
		C & 96.5\% & 94.0\% & 95.2\% & 91.4\% \\
		D & 93.8\% & 98.1\% & 95.9\% & 93.0\% \\
		E & 95.3\% & 95.3\% & 95.3\% & 92.2\% \\
		F & 94.4\% & 91.8\% & 93.1\% & 88.3\% \\
		\hline
	\end{tabular}
\end{table}

\subsubsection{Tensorflow for Poets}
\label{sec:TF_for_poets}

\textquoteleft Tensorflow for Poets' uses the \textquoteleft Inception v3' network, a pre-trained deep neural network that is trained for ImageNet Large Visual Recognition Challenge. It is able to differentiate between 1000 different classes. We used this approach to perform classifications and compare the results to the custom-designed networks using the Lasagne library, however we found the results to be inferior. This poorer performance can be explained by the fact that the class types trained on are mainly examples of every-day objects and animals rather than scientific images. Another reason is that using a custom-designed network has much more freedom in adjusting parameters compared to using a \textquoteleft black-box' approach, where more parameters are fixed.

\section{Results for four classes}
\label{sec:results_3}

In the previous section we have explored varying several parameters using the custom designed network in Lasagne and found the optimal one that results in the highest test classification accuracy for the two-class problem of distinguishing between compact and multiple component extended sources, which was the 3 convolutional and 2 dense layer architecture without sigma clipping (model E), using both original and augmented images. Given these results, we wanted to see how well such a deep neural network setup could distinguish between two additional classes of data, consisting of single component extended and two component extended sources. 

The same parameters were used as was described in Section~\ref{sec:DL_algorithms}. Two models were explored for the task of four-class classification; the 3 convolutional and 2 dense layer architecture with and without sigma clipping, using both original and augmented images. This architecture and set of images performed best on the two-class problem, which is why it was chosen for the four-class problem. The numbers of images used are summarised in Table \ref{tab:Four_class_problem1}. The issue with class imbalance was addressed by augmenting the single, two and multiple-component images to achieve roughly the same number of images for these extended sources. We used the same set of original and augmented images for the compact sources as for the two-class problem. The results shown in Table~\ref{tab:Results4} are the classifier metrics on the validation and test data sets, as was done similarly for the two-class problem, however applying a \textquoteleft macro' average over the four classes to obtain an overall summary of the number of true and false positives and negatives across the confusion matrix.

\begin{table}
	\centering
	\caption{Results for four-class model. The difference between using sigma clipping or not is very minor, and can be attributed to random fluctuations for each subsequent run.}
	\label{tab:Results4}
	\begin{tabular}{lcccr}
		\hline
		& \textbf{Valid.} Precision & Recall & F1 & Accuracy \\
		\hline
		E & 92.6\% & 92.7\% & 92.7\% & 92.0\%  \\
		F & 93.2\% & 93.3\% & 93.2\% & 92.7\% \\
		\hline
		& \textbf{Test} Precision & Recall & F1 & Accuracy \\
		\hline
		E & 94.0\% & 94.1\% & 94.0\% & 93.5\%  \\
		F & 94.0\% & 93.9\% & 93.9\% & 93.5\% \\
		\hline
	\end{tabular}
\end{table}

\begin{table}
	\centering
	\caption{Individual precision and recall values computed from the confusion matrix for the 4-class test set, using the original and augmented images, for model E.}
	\label{tab:prec_rec_4_class}
	\begin{tabular}{lcr}
		\hline
		 & Precision & Recall \\
		\hline
		Compact & 96.9\% & 97.4\%  \\
		\hline
		Single-extended & 93.4\% & 95.3\% \\
		\hline
		Two-component extended & 91.1\% & 87.6\% \\
		\hline
		Multiple-component extended & 94.6\% & 96.1\%  \\
		\hline
	\end{tabular}
\end{table}

\begin{figure}
    \centering
    \includegraphics[width=90mm,scale=0.9]{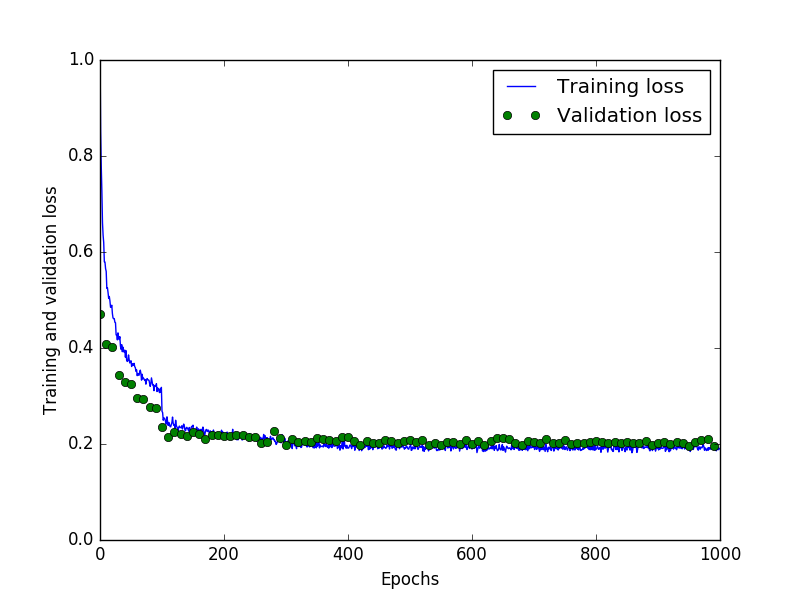}
    \caption{Training and validation losses shown when using a chunk size of 20000 with a 3 conv + 2 dense model with no sigma clipping (model E), for the four-class problem. The training losses are much higher at the start compared to what was observed in the 2-class problem, and settle to a loss of around 0.2 by 400 training epochs.}
    \label{fig:four_class_plots_combined}
\end{figure}

The inclusion of an additional two classes of data results in a significantly reduced performance compared to when only two classes are used. This is likely due to the low-level amount of label contamination in the new classes, in addition to the level already present in the previous two classes. The manually chosen images for the multiple-component extended source class still contain a substantial number of images that are superpositions. Despite this, the accuracies remain above 93\%. 

Note that our machine learning algorithms are still making the correct decision in determining membership in each of the four classes, in that they were trained to simply recognise the number of extended components in close proximity. The information required to identify some of these images as superpositions of more than one physical radio galaxy requires more detailed information about both the radio morphology and the location of possible optical/IR counterparts. This is a future task for machine learning algorithms to make use of labels with higher-level information, for example from Radio Galaxy Zoo.

The individual precision and recall values were computed for each of the four classes in Table~\ref{tab:prec_rec_4_class}. The results show that the precision and recall values are the highest for the compact sources, so the deep learning algorithm is able to identify all the compact sources and not confuse them with any other source, with most accuracy. This is to be expected since they have a very well-defined morphology with the least variability amongst the classes. The deep learning algorithm however produces the lowest scores for the two-component extended sources, and this is likely because there is the most overlap between these sources and the two classes on either side; the single-extended and multiple-extended sources. 

These higher-level classes of data can be used as an initial step to facilitate the generation of more specific radio morphology classes of scientific interest.

\subsection{Comparing results with Data Release 1 of the Radio Galaxy Zoo}
\label{sec:DR1}

Radio Galaxy Zoo classification is a two-step process. For a single classification, users firstly select all radio emission they consider to be originating from a single radio galaxy. After selecting the radio components, users will try to match it with a galaxy in the near-IR data. If there are multiple radio sources in the image, users can repeat both steps to identify other radio galaxies. Individual classifications are aggregated to provide a consensus classification of the image based on the majority vote \citep{Willett_2016}.

Data Release 1 of the Radio Galaxy Zoo (DR1; Wong et al. in preparation) was made with the purpose of obtaining citizen-scientists input in identifying which components belonged together in a source. The \textquoteleft Number of components' is defined as the number of discrete radio components that a source encompasses, that also depends on the lowest radio contour level chosen. The \textquoteleft Number of peaks' that examines the components identified by RGZ participants, refers to how many peaks are present in the radio source as determined by an automatic pipeline processor. DR1 consists of 74627 sources where user weightings have been applied to the consensus levels, retaining the sources which have a consensus level of 0.65 or higher. The minimum reliability of DR1 is 75\% for a minimum weighted consensus level of 0.65 for the classifications of the FIRST survey. Using the dataset for the four-class problem consisting of 21933 images, there are 10722 (14.4\%) images in common with the DR1 dataset, where the matching is done based on the source name. After removing the sources that contained invalid entries in the \textquoteleft matchComponents'  and \textquoteleft WISECATmismatch' columns, there were 9537 remaining (12.8\%). 

Using the \textquoteleft Number of components' and \textquoteleft Number of peaks' information provided that originated from the citizen scientists' and the post-processing pipeline, along with the images in the DR1 dataset, we were able to generate labels for the overlapping dataset of 9537 images. Since there is no way of distinguishing between compact sources and extended sources based on this information alone, we decided to make a single class composed of compact and single-component extended images. The labels for the classes were generated using the rules as shown in Table~\ref{tab:DR1_labels}. These sources make up the test set, to assess how well the custom designed network in Lasagne is able to reproduce the labels generated based on the citizen-scientists input. The sources where no class could be assigned were removed, leaving 6966 sources. The remaining images that were not part of the test set of intersected images formed the training and validation set. These numbers are summarised in Table~\ref{tab:point_extended_proportions}. It is worth noting that the original set of images again contains an imbalance in the number of sources belonging to each class, where there are fewer compact/single-extended sources and the fewest multiple-component extended sources. This imbalance is compensated by augmenting these classes more. 

The architecture used is the three convolutional and two dense layer architecture since this is the overall best-performing architecture. Two datasets are used; the first one using just the original images that contain imbalanced classes, as well as the original and augmented images that contain much more balanced numbers of images in the classes. The parameters used for these two datasets are summarised in Table~\ref{tab:Parameters_DR1}. 

The results show that when using just the original images, the precision and recall metrics are quite low overall, as shown in the first 
row of Table~\ref{tab:DR1_results}. Upon exploring the individual metrics for the three classes in 
Table~\ref{tab:prec_rec_DR1_orig}, the deep learning algorithm is able to identify the 
compact/single-extended sources effectively, however it struggles more with identifying the two-component extended sources, 
despite there being more examples of this class to train on. The most likely reason is that the DR1 data contains more information for each source compared to what the deep learning algorithm is trained on. Based on the input from the citizen scientists, it will take a 2 component extended source and divide it into two 1 component sources, depending on the WISE ID status. The deep learning algorithm performs exceptionally poorly with the multiple-component extended images, which is not surprising given that there are only several hundred examples of this class of images to train and validate on. 

In using the augmented images that have been generated to even out the class imbalance, in addition to the original
images, all the average metrics are improved, as can be seen in the second row of Table~\ref{tab:DR1_results}. Upon examining the 
individual metrics for each class in Table~\ref{tab:prec_rec_DR1_orig_aug}, the precision values are improved across all the classes. The recall values are improved for the compact/single-extended class, and are substantially higher for the multiple-component extended class compared to when only the original images are used, however they still remain quite low for 
this class. The deep learning algorithm is therefore much less precise and sensitive in identifying the images belonging to the multiple-component extended class, when the labels are generated according to citizen 
scientists input, compared to the other two classes. It does not perform as well in detecting the images that are labelled as multiple-component extended sources, and it also predicts images as being in this class when they are labelled as belonging to another class. A couple of reasons are as follows. There were only on the order of a few hundred (475) original images to train on for images in the multiple-component extended source class, and although they are augmented to generate a set of images that has a roughly the same number compared to the other classes, there are perhaps not enough original examples of the different morphologies that can exist, therefore making the feature space smaller for this class. Additionally, although the multiple-component extended sources in the training and validation set were inspected in an attempt to ensure that the images contain at least three components that are part of the same source, which was the classification scheme used by RGZ users, there were still found to be a substantial number of images that contained source superpositions, upon cross-checking with several optical/IR images. However it is important to keep in mind that the deep learning algorithm was trained to recognise the number of extended components in close proximity, using radio galaxy images only. It should be noted that all multi-component sources, whether they are superpositions or not, belong in the multi-component class.

Presumably, the higher the number of components an image appears to contain, the more likely it is that the images are superpositions of sources. This would explain why the compact/single-extended and two-component extended sources are not affected as much in terms of the precision and recall metrics as the multiple-component extended class. It should further be noted that 77.6\% of images belong to the compact/single-component extended class, which explains the overall high classification accuracies in Table~\ref{tab:DR1_results}.

The generation of augmented images to even out the imbalance in classes in the original data overall improves the metrics in predicting the labels that are generated using citizen-scientists input.

\begin{table}
	\centering
	\caption{Rules by which labels were generated for the DR1 dataset, based on citizen scientists input, to test the best-performing Lasagne convolutional neural network architecture. The number given refers to both the number of components and number of peaks in a given image. For example, the Compact/Single-extended class is defined as having 1 component and 1 peak.}
	\label{tab:DR1_labels}
	\begin{tabular}{lr}
		\hline
		\# components and \# peaks & Label \\
		\hline
		1 & Compact/Single-extended \\
		\hline
		2 & Two-component extended \\
		\hline
		$\geq$ 3 & Multiple-component extended \\
	\end{tabular}
\end{table}

\begin{table}
	\centering
	\caption{Summary of the numbers of sources used for training, validation and testing of the labels generated from the DR1 data, for both the original (Orig.) and augmented (Aug.) images.}
	\label{tab:point_extended_proportions}
	\begin{tabular}{lcc}
		\hline
		Data & \# Orig. & \# Orig. + Aug.  \\
		\hline
		DR1 & 74627 & \\
		Final intersected dataset (Test) & 6966 \\
		Compact/Single-extended (Train) & 4147 &14588\\
		Two-component extended (") & 10306 &14306\\
		Multiple-component extended (") & 475 & 15177\\
		\hline
	\end{tabular}
\end{table}

\begin{table}
	\centering
	\caption{Chunk sizes and percentage of data used for training, validation and testing for the Lasagne deep learning network in the DR1 cross-check analysis.}
	\label{tab:Parameters_DR1}
	\begin{tabular}{lcccr}
		\hline
		 & Chunk size & \% Train. & \% Valid. & \% Test  \\
		\hline
		Orig. & 1000 & 59\% & 9\% & 32\%  \\
		Orig.+Aug. & 3000 & 78\% & 8\% & 14\%  \\
		\hline
	\end{tabular}
\end{table}

\begin{table}
	\centering
	\caption{Validation and Test metrics for the DR1 cross-check analysis.}
	\label{tab:DR1_results}
	\begin{tabular}{lcccr}
		\hline
		 & \textbf{Valid.} Precision & Recall & F1 & Accuracy \\
		\hline
		Orig. & 89.7\% & 58.7\% & 58.2\% & 86.4\% \\
		Orig.+Aug. & 90.6\% & 90.6\% & 90.5\% & 90.7\% \\
		\hline
		 & \textbf{Test} Precision & Recall & F1 & Accuracy \\
		\hline
		Orig. & 75.6\% & 62.6\% & 61.6\% & 92.8\% \\
		Orig.+Aug. & 79.6\% & 81.6\% & 80.6\% & 94.8\% \\
		\hline
	\end{tabular}
\end{table}

\begin{table}
	\centering
	\caption{Individual precision and recall values computed from the confusion matrix for the DR1 test set of 6966 images, when training on just the original images.}
	\label{tab:prec_rec_DR1_orig}
	\begin{tabular}{lcr}
		\hline
		 & Precision  & Recall \\
		\hline
		Compact/Single-extended & 97.2\% & 95.0\%  \\
		\hline
		Two-component extended & 79.5\% & 90.7\% \\
		\hline
		Multiple-component extended & 50.0\% & 2.1\%  \\
		\hline
	\end{tabular}
\end{table}

\begin{table}
	\centering
	\caption{Individual precision and recall values computed from the confusion matrix for the DR1 test set of 6966 images, when training on both the original and augmented images.}
	\label{tab:prec_rec_DR1_orig_aug}
	\begin{tabular}{lcr}
		\hline
		 & Precision  & Recall \\
		\hline
		Compact/Single-extended & 97.5\% & 96.9\%  \\
		\hline
		Two-component extended & 88.0\% & 89.5\% \\
		\hline
		Multiple-component extended & 53.4\% & 58.5\%  \\
		\hline
	\end{tabular}
\end{table}

\section{Conclusions}
\label{sec:conclusions}

This is a methods paper that explored the use of deep neural networks for classifying compact and various classes of extended sources in radio astronomical data. We have found an optimal set of parameters obtained
from examining the two-class problem of distinguishing between two well-defined classes of data composed of compact and multiple-component extended sources, and applied this to a classification scenario involving more classes, and have shown that the classification accuracies remain high without excessive overfitting. The results were cross-checked on the Radio Galaxy Zoo DR1 dataset, where the generation of augmented images in order to address the class imbalance highly influenced the accuracies to predict the labels generated based on the citizen scientists input. However, the predictions for the multiple-component extended class remained poor, most likely because this dataset contained the fewest number of original images to train on, and did not have the additional information of which components made up a radio source and how many peaks were contained in the source, which was the additional information provided in the DR1 dataset.

The first part of the results explored various architectures and identified the optimal parameters for distinguishing between the two morphological extremes of compact and multiple-component extended sources. We found that the three convolutional and two dense layer architecture using the original and augmented images with no sigma clipping produced the maximal accuracy of 97.4\% for the two-class problem, which is significantly better compared to using just the original images with the same architecture. Although the equivalent architecture with sigma-clipping produced an accuracy in the same range, the difference between the training and validation loss was greater. A better model is ensured if the training and validation losses are closer together. The largest influence of performance other than the model architecture was to use a relatively large chunk size, since the more examples that are seen simultaneously, the better the estimate can be for adjusting the weights to achieve a lowered cost function. This is where the use of augmented data is useful, as it allows one to use a larger chunk size. Another important impact on the performance of the deep neural network is to use quite a small learning rate at the start and make it smaller by a factor of 10 at certain points during training, and using a small batch size of 8 samples.   

When training deep neural networks with a large enough number of images, removing noise through the use of sigma clipping appears to offer no significant benefit. Given there is an adequate number of images belonging to the available classes in question, with varying levels of noise, the deep learning network can learn these properties and become robust to them. 

Using the knowledge gained from the factors that influence the performance of the classifier in the two-class problem, we assumed that the setup would perform similarly for distinguishing between an additional two classes of images. It is unclear what the effect would have been, had two classes been chosen that were not extreme examples of morphologies. For the four-class problem of distinguishing between compact, single, two-component and multiple-component extended sources, and using the three convolutional and two dense layer setup with original and augmented images, we were able to achieve a classification accuracy of 93.5\%. The fact that the compact and single-component extended sources are both chosen from where PyBDSF has detected one component, and that the deep learning algorithm is able to achieve high precision and recall values for these two classes, means that the deep learning algorithm is doing more that just counting the number of components in the images. 

Both the two-class and four-class problems contain different numbers of original images in each class. This did not appear to dramatically affect the performance of the classifier when using the original set of images in the two-class problem, most likely because the minority set of images was comprised of compact sources that have a very specific morphology, and the sources are almost always found the in the centre of the image. 

It is worth noting that at least 44\% of images in the multiple-component extended class in the two-class problem appeared to contain superpositions, or fewer than three components. Although we attempted to remove these images in the four-class problem by manually selecting the sources, a substantial number of images with superpositions remained, upon cross-checking with several optical/IR images. However, the deep learning algorithm was trained to identify extended components in close proximity in a radio galaxy image, so it is still making the correct decisions in determining class membership based on using the image data alone.

The other classes explored apart from compact sources display a much richer variety of morphologies, which is why it is important to augment those images much more in comparison to the compact sources. Roughly equal augmented datasets were generated for the extended source classes in the four-class problem, to make up for the class imbalance present in the original images. This was especially important for the DR1 analysis, where the deep learning algorithm was much better able to predict the labels generated based on citizen scientists input when the augmented data was used in addition to the original data, to compensate for uneven classes. Although the precision and recall values for the compact/single-component extended sources is quite high, it is possible to use linear regression and simple positional matches to identify such sources. The metrics were moderately high for the two-component extended sources. The deep learning algorithm however struggles more to identify the multiple-component sources when the labels are generated using input from the citizen-scientists, as is evidenced from the poorer precision and recall values for this class of images. This indicates the need for both more original images and labels with higher-level information from citizen scientists to make up the training and validation set, in order to predict these sources more accurately. The value in using both data from the RGZ as well as the help of computer algorithms is the ability to connect discrete individual components that may be associated with a source.

The first example of using convolutional neural networks to classify radio morphologies was in \citet{Aniyan_Thorat2017}, where they choose a couple of hundred examples of FRI, FRII and Bent-tailed galaxy morphologies, perform sigma clipping, apply a high amount of augmentation, and build a fusion classifier to combine the results for the three classes. However, the authors run into problems of substantial overfitting, due to not using enough examples of different varieties of FRI, FRII and Bent-tailed classes. An earlier study using an unsupervised learning approach consisting of Kohonen maps has shown that when categorising radio galaxies into FRI and FRII type sources, sigma clipping and other pre-processing may be necessary \citep{Polsterer_etal2016}. In contrast, the current work has shown that with enough examples of broad classes of radio galaxy morphologies, it appears that pre-processing and noise removal through sigma clipping does not offer a significant advantage and that it is possible to classify radio galaxy morphologies into more than two classes using only convolutional networks, without a high level of overfitting.

The use of deep learning networks appears to be very well suited to source classification in radio surveys. However one must keep in mind that the deep learning algorithm will only able to make predictions that are as good as the level or complexity of information that is input into it. When there are a limited number of people to make the classifications, one option to sift through the vast amount of data is to use automated techniques such as PyBDSF or blob-detection algorithms, to assist in providing structure. However these techniques do not always reflect how humans would classify images; they are poorer at making the distinction between images containing superpositions, and images containing sources that have multiple components associated with each other. They can also detect components that a human would identify as noise, as shown in Fig.~\ref{fig:3_component_image}. Therefore it is more likely that there will be contaminations in the training set. However, given access to the classifications from an increasing consensus of people that are trained to identify which components belong together in a particular image, the training labels will be more accurate, as will the predictions. Citizen Science projects like RGZ are an excellent way of generating training sets, and appear to have a reliability similar to that of trained astronomers.

When there are few people available to make classifications, there are limitations in the extent of human intervention that can be applied to reduce contamination in the data. In this case, the results shown indicate that it is better to devote more time in further classifying the images where PyBDSF has detected only up to a few components, as they are less likely to contain superpositions. 

The labels generated with the help of algorithms such as PyBDSF are able to attain a certain level of concordance when compared to labels used from citizen-scientists. However, they appear unable yet to replace input from humans, who are able to detect finer-scale structures and subtle aspects of morphologies such as the amount and direction in which the bulges in the edges of radio components are pointing and how far apart they are, that influences whether the components are associated with each other, for a source in question. With the availability of higher-level training labels provided by humans as opposed to the lower-level ones provided by automated techniques such as PyBDSF, deep-learning techniques should exceed the performance of PyBDSF in the future.

Another consideration is the identification of rare sources such as radio relics that make up a small fraction of the overall observed morphologies. Although they are more likely to be found in those images where PyBDSF has detected a multitude of components, these images contain an increasing number of source superpositions, so it is still necessary to have humans to visually inspect the source to see whether they are true relics or not, since PyBDSF has certain ways of grouping the gaussians that are fit to the sources, that may not match how a person would associate them, even when changing parameters that control how the components are grouped. 

In future work, we aim to optimise deep neural network setups for more complex morphological classifications and will use them on LOFAR survey data (LOTSS, \citet{Shimwell_etal2016}). We will also explore neural networks that perform cross-identification with optical/IR surveys \citep{Norris_2016}.

\section*{Acknowledgements}

We thank Peter Schleper and Christoph Garbers for helpful discussions on the topic. We thank Chris Snyder, Ed Paget, Enno Middelberg, Kyle Willett, Heinz Andernach, Rob Simpson, Amit Kapadia, Arfon Smith, Melanie Gendre and Samuel George for their contribution to RGZ. We also acknowledge C. Synder, E. Paget, E. Middelberg, R. Simpson, A. Kapadia, A. Smith, M. Gendre, and S. George who have made contributions to the project from the beginning and have since moved on. We thank Matthew Alger, Stas Shabala, Jean Tate and Laurence Park for helpful comments on earlier drafts of the paper.

VL acknowledges support by the Deutsche Forschungsgemeinschaft (DFG) through grant SFB 676.

Partial support for this work at the University of Minnesota comes from grants AST-1211595 and AST-1714205 from the U.S. National Science Foundation. The FIRST survey was conducted on the Very Large Array of the National Radio Astronomy Observatory, which is a facility of the National Science Foundation operated under cooperative agreement by Associated Universities, Inc.

BDS gratefully acknowledges support from Balliol College, Oxford through the Henry Skynner Junior Research Fellowship and from the National Aeronautics and Space Administration (NASA) through Einstein Postdoctoral Fellowship Award Number PF5-160143 issued by the Chandra X-ray Observatory Center, which is operated by the Smithsonian Astrophysical Observatory for and on behalf of NASA under contract NAS8-03060.

This publication has been made possible by the participation of more then 10,000 volunteers in the Radio Galaxy Zoo project. The data in this paper are the results of the efforts of the Radio Galaxy Zoo volunteers, without whom none of this work would be possible. Their efforts are individually acknowledged at http://rgzauthors.galaxyzoo.org. J.K.B. acknowledges financial support from the Australian Research Council Centre of Excellence for All-sky Astrophysics (CAASTRO), through project number CE110001020.












\bsp	
\label{lastpage}
\end{document}